\documentstyle[12pt]{article}

\def\nsection#1{\section{#1}\setcounter{equation}{0}}

\advance \textheight by \topskip
\begin{document}

\author{Mikhail S. Plyushchay\thanks{On leave from the
{\it Insti\-tu\-te for High Ener\-gy Phy\-sics, Protvino, Russia;}
E-mail: mikhail@posta.unizar.es}\\ \\
{\it Departamento de F\'{i}sica Te\'orica, Facultad de Ciencias}\\
{\it Universidad de Zaragoza, 50009 Zaragoza, Spain}}

\title{\bf Minimal bosonization of supersymmetry}

\date{To appear in {\bf Mod. Phys. Lett. A}}

\maketitle

\begin{abstract}
The minimal bosonization of supersymmetry in terms of one bosonic
degree of freedom is considered.  A
nontrivial relationship of the construction to the Witten
supersymmetric quantum mechanics is illustrated with the help of
the simplest $N=2$ SUSY system realized on the basis of the
ordinary (undeformed) bosonic oscillator.  It is shown that the
generalization of such a construction to the case of
Vasiliev deformed bosonic oscillator gives a supersymmetric
extension of the 2-body Calogero model in the phase of exact or
spontaneously broken $N=2$ SUSY. The construction admits an
extension to the case of the OSp(2$\vert$2) supersymmetry, and,
as a consequence, $osp(2\vert 2)$ superalgebra is revealed as a
dynamical symmetry algebra for the bosonized supersymmetric
Calogero model.  Realizing the Klein operator as a parity
operator, we construct the bosonized Witten supersymmetric
quantum mechanics.  Here the general case of the corresponding
bosonized $N=2$ SUSY is given by an odd function being a
superpotential.
\end{abstract}
\vskip1.5cm
\begin{flushright}
{\bf DFTUZ/95/09}\\
{\bf hep-th/9601141}
\end{flushright}
\newpage

\nsection{Introduction}

The possibility of describing (1+1)-dimensional fermionic
systems in terms of bosonic fields
is known for a long time \cite{klei,bosf}. The corresponding
Bose-Fermi transformation is the Klein transformation \cite{klei,bos1},
which has a nonlocal nature. Such a nonlocality lies also in
the basis of (2+1)-dimensional anyonic constructions \cite{any}.
Analogous Bose-Fermi transformation (bosonization) exists in the
(0+1)-dimensional case
of quantum mechanics \cite{bos1,garb}, and its generalization
leads to the $q$-deformed oscillator \cite{bosq}.

It is obvious that the bosonization constructions can be
straightforwardly generalized to the case of supersymmetric
quantum mechanics\footnote{In the last section we shall comment
on the attempt of applying bosonization technique to
(1+1)-dimensional SUSY systems \cite{damg}.}. Indeed, realizing
fermionic oscillator operators in terms of creation-annihilation
bosonic operators and extending the bosonized fermionic system
by independent bosonic oscillator operators, we can realize
$N=2$ supersymmetric system following the Nicolai-Witten
supersymmetric quantum mechanical constructions \cite{nic,wit}.
But, as we shall see,  such a straightforward construction turns
out to be a {\it nonminimal} one.

In this paper, we investigate the possibility of realizing
supersymmetric quantum mechanical bosonization constructions in
a {\it minimal way}, in terms of one bosonic degree of freedom
in the simplest case.  As we shall see, the crucial difference
of the minimal bosonization scheme from the nonminimal one is
coded in relation (\ref{af}).

For the first time the  possibility of revealing superalgebraic
structures in a quantum system of one bosonic oscillator was pointed out,
probably, in ref. \cite{cromb}, where it was noted that the
$osp(1\vert 2)$ superalgebra is the spectrum generating algebra
of the system. Explicit realization of this algebra
in terms of creation-annihilation bosonic operators was given in
ref. \cite{macmaj}, and subsequently generalized in ref.
\cite{mac} to the case of Vasiliev deformed bosonic
oscillator.  This deformed bosonic oscillator was introduced in
ref. \cite{vas1} in the context of higher spin algebras, and
subsequently was used for investigation of the quantum
mechanical Calogero model \cite{bri1}--\cite{bri2}  and for
constructing (2+1)-dimensional anyonic field equations
\cite{ply1}.

The deformed Heisenberg algebra, corresponding to the deformed
bosonic oscillator of ref. \cite{vas1}, involves the Klein
operator as an essential object, which introduces $Z_2$-grading
structure on the Fock space of the system. Such a structure, in
turn, is an essential ingredient of the $N=2$ supersymmetry,
which was interpreted in ref. \cite{mac} as a {\it hidden
supersymmetry} of the {\it deformed} bosonic system.

In the present paper we shall demonstrate that a `hidden' $N=2$ supersymmetry,
revealed in ref. \cite{mac}, has a nonlocal nature analogous to
that of the standard bosonization constructions for fermionic
systems \cite{klei,bosf,bos1,garb}.  We shall show that the
simplest $N=2$ SUSY system, constructed on the basis of the
deformed Heisenberg algebra, is the bosonized $N=2$
supersymmetric 2-body Calogero model, and that the bosonization
constructions of $N=2$ SUSY can be generalized to the case of
OSp(2$\vert$2) supersymmetry.  Moreover, it will be shown how
the simplest bosonization constructions can be generalized to
the case corresponding (formally) to the Witten supersymmetric
quantum mechanics.  This generalization will allow us to
demonstrate that the bosonized supersymmetric quantum mechanics
constructed on the basis of the deformed Heisenberg algebra
gives the same results as the general supersymmetric bosonization
scheme realized on the basis of the Heisenberg algebra of ordinary,
undeformed, bosonic oscillator.

The paper is organized as follows.  In section 2 we realize the
simplest $N=2$ SUSY system with the help of the Heisenberg
algebra supplied with the Klein operator which, in the
coordinate representation, can be considered as a parity
operator.  We show that such a construction contains both phases
of exact and spontaneously broken SUSY \cite{wit,alv}.  Here
these two phases are distinguished by the sign parameter being
present in the bosonized $N=2$ SUSY generators.  We trace out a
formal analogy between this simplest bosonized SUSY system and
the simplest system of the Witten supersymmetric quantum
mechanics \cite{wit}.  The latter one is the Nicolai superoscillator
\cite{nic}, which, in contrast to the constructed system, contains
only the phase of unbroken SUSY.  Proceeding from the phase of
the broken SUSY, we construct fermionic creation-annihilation
operators, i.e. realize a Bose-Fermi (Klein) transformation.
This allows us to reveal the point where the minimal
bosonization of supersymmetry essentially differs from the nonminimal
one.

In section 3 we generalize the constructions to the case of
deformed bosonic oscillator \cite{vas1}, whose
creation-annihilation operators satisfy the deformed Heisenberg
algebra involving the Klein operator.  The specific feature of
such a generalization is that in the phase of spontaneously
broken supersymmetry, the scale of supersymmetry breaking is
governed by the deformation parameter of the system.  On the
other hand, the phase of exact supersymmetry turns out to be
isospectral to the corresponding phase of the simplest bosonized
$N=2$ SUSY system from section 2.  We show that this generalized
construction, connected with the deformed Heisenberg algebra,
gives the supersymmetric extension of the 2-body Calogero model
\cite{calog} realized, in contrast to the standard approach
\cite{freed,bri2}, without extending the system by the fermionic
degrees of freedom.

In section 4 we demonstrate that the construction admits an
extension to the case of the OSp(2$\vert$2) supersymmetry.  This
means, in particular, that the corresponding $osp(2\vert 2)$
superalgebra is realizable as an operator algebra for the quantum
mechanical 2-body nonsupersymmetric Calogero model. On
the other hand, it is the algebra of a dynamical symmetry for the
bosonized supersymmetric extension of the 2-body Calogero model
presented in section 3.  Thus, we bosonize all the constructions of
refs. \cite{freed,bri2} corresponding to the case of the
2-body supersymmetric  Calogero model.

In section 5 we generalize the $N=2$ SUSY bosonization
constructions to the case which corresponds (formally) to the
general case of the Witten supersymmetric quantum mechanics.
Here we show that from the point of view of
the bosonized supersymmetric constructions, the deformed
Heisenberg algebra gives the same general results as the
undeformed Heisenberg algebra.  A priory this fact is not to be
surprising, since the bosonization scheme is based on the use of
the Klein operator, and from the point of view of its
realization as a parity operator, it is exactly one and the same
object for both cases.

In section 6 we list some problems which may be interesting
for further consideration.

\nsection{The simplest $N=2$ SUSY system}
We shall begin with the construction of the simplest bosonized supersymmetric
system. This will give the basis for subsequent generalizations
and will allow us to demonstrate the nontrivial relationship of the bosonized
supersymmetry to the standard supersymmetry realized by
supplying the bosonic system with independent fermionic operators
\cite{wit,salhol}.

So, let us consider the ordinary bosonic oscillator
with the operators $a^{+}$ and $a^{-}$ satisfying
the commutation relation
$
[a^{-},a^{+}]=1,
$
and introduce the Klein operator $K$ defined by the relations
\begin{equation}
K^{2}=1,\quad \{K,a^{\pm}\}=0.
\label{kle}
\end{equation}
This operator separates the complete orthonormal set of states
$|n>=(n!)^{-1/2}(a^+)^n \vert 0>$, $n=0,1,\ldots$, $a^-\vert 0>=0$,
into even and odd subspaces:
$
K|n>=\kappa\cdot (-1)^{n}|n>,
\label{z2}
$
and, so, it introduces $Z_2$-grading structure on the Fock space of
the bosonic oscillator. Without loss of generality,
we fix the sign factor as $\kappa=+1$.
The operator $K$ can be realized as
$
K=\exp{i\pi N},
$
or in the explicitly hermitian form,
\begin{equation}
K=\cos \pi N,
\label{kcos}
\end{equation}
with the help of the number operator $N=a^{+}a^{-}$,
$
N|n>=n|n>.
$
Before going over to the SUSY
constructions, we note that in the
coordinate representation,
where the creation-annihilation operators
are realized as $a^{\pm}=(x\mp ip)/\sqrt{2},$
$p=-id/dx$, the Klein operator can be considered
as the parity operator, whose action is defined by the relation
$K\psi(x)=\psi(-x)$, and, so, the space of wave functions
is separated into even and odd subspaces,
\begin{equation}
K\psi_{\pm}=\pm\psi_\pm(x),\quad \psi_\pm(x)=
\frac{1}{2}(\psi(x)\pm\psi(-x)),
\label{kpsi}
\end{equation}
in correspondence with relations (\ref{kle}) and the
above mentioned choice of the sign parameter $\kappa$.  It is
the relations (\ref{kpsi}) that we shall consider as defining
the Klein operator in the coordinate representation.  But, on
the other hand, if here we write the exact analog of the
expression (\ref{kcos}),
$
K=\sin (\pi H_0),
$
$
H_{0}=\frac{1}{2}(x^{2}-{d^{2}}/{dx^{2}}),
$
we shall
immediately reveal the hidden nonlocal character of the
bosonization constructions.

Now let us proceed to the supersymmetric constructions.
For realizing $N=2$ supersymmetry, we shall construct the
mutually conjugate nilpotent operators $Q^{+}$ and
$Q^{-}=(Q^{+})^{\dagger}$, $Q^{\pm 2}=0$.
We shall look for the simplest possible realization of such operators
in the form
$
Q^{+}=\frac{1}{2}a^{+}(\alpha+\beta K)+\frac{1}{2}a^{-}(\gamma+\delta K),
$
which is linear in the oscillator variables
$a^{\pm}$ but contains also the dependence on the Klein operator $K$.
Then, the nilpotency condition,
$
Q^{+2}=Q^{-2}=0,
$
gives the following restriction on the complex number parameters:
$
\beta=\epsilon\alpha,\quad \delta=\epsilon\gamma,
$
where $\epsilon=\pm$.
Therefore, we have two possibilities for choosing  operator $Q^{+}$:
$
Q^{+}_{\epsilon}=(\alpha a^{+}+\gamma a^{-})\Pi_{\epsilon},$
which are distinguished by the sign parameter.
Here we introduced a notation $\Pi_{\epsilon}$
for hermitian operators
$
\Pi_{\pm}=\frac{1}{2}(1\pm K)
$
being the projectors:
$\Pi_{\pm}^{2}=\Pi_{\pm},$ $\Pi_{+}\Pi_{-}=0,$ $\Pi_{+}+\Pi_{-}=1.$
{}From the explicit form of the anticommutator,
$
\{Q^{+}_{\epsilon},Q^{-}_{\epsilon}\}=
a^{+2}\alpha\gamma^{*}+a^{-2}\alpha^{*}\gamma
+\frac{1}{2}\{a^{+},a^{-}\}(\gamma\gamma^{*}+\alpha\alpha^{*})
-\frac{1}{2}\epsilon K
[a^{-},a^{+}]
(\gamma\gamma^{*}-\alpha\alpha^{*}),
$
we conclude that it will commute with the number
operator $N$ if we choose the parameters in such a way that
$\alpha\gamma^{*}=0$. As a consequence, in this case
the spectra of the corresponding Hamiltonians,
$H_\epsilon=\{Q^{+}_{\epsilon},Q^{-}_{\epsilon}\}$, $\epsilon=\pm$,
will be the simplest, linear in $n$.
We can put $\alpha=0$ and normalize the second parameter as
$\gamma\gamma^*=1$.
The remaining phase factor can be removed by the unitary transformation
of the oscillator operators $a^{\pm}$, and, so,
finally we arrive at the nilpotent
operators in the compact form:
\begin{equation}
Q^{+}_{\epsilon}=a^{-}\Pi_{\epsilon},\quad
Q^{-}_{\epsilon}=a^{+}\Pi_{-\epsilon}.
\label{qexp}
\end{equation}
They together with the operator
\begin{equation}
H_{\epsilon}=\frac{1}{2}\{a^{+},a^{-}\}
-\frac{1}{2}\epsilon K [a^{-},a^{+}]
\label{hexp1}
\end{equation}
form the $N=2$ (or $s(2)$, according to the terminology of ref.
\cite{cromb}) superalgebra:
\begin{equation}
Q^{\pm2}_{\epsilon}=0,\quad \{Q^{+}_{\epsilon},Q^{-}_{\epsilon}\}
=H_{\epsilon},\quad
[Q^{\pm}_{\epsilon},H_{\epsilon}]=0.
\label{salg}
\end{equation}
Note that the hermitian supercharge operators
$Q^{1,2}_{\epsilon}$,
$
Q^{\pm}_{\epsilon}=\frac{1}{2}(Q^{1}_{\epsilon}\pm iQ^{2}_{\epsilon}),
$
$
\{Q^{i}_{\epsilon},Q^{j}_{\epsilon}\}=2\delta^{ij}H_{\epsilon},
$
have the following form in terms of coordinate and momentum operators:
\begin{equation}
Q^{1}_{\epsilon}=\frac{1}{\sqrt{2}}(x+i\epsilon pK),\quad
Q^{2}_{\epsilon}
=\frac{1}{\sqrt{2}}(p-i\epsilon xK)=-i\epsilon Q^{1}_{\epsilon}K.
\label{q12}
\end{equation}

Let us consider the spectrum of the constructed
SUSY Hamiltonian (\ref{hexp1}). In the case when $\epsilon=-$,
the states $|n>$ are the eigenstates  of the operator $H_{-}$ with
the eigenvalues
\begin{equation}
E_{n}^{-}=2[n/2]+1,
\label{eb}
\end{equation}
where $[n/2]$ means the integer part of $n/2$. Therefore, here
$E_{n}^{-}>0$ and all the states
$|n>$ and $|n+1>$, $n=2k$, $k=0,1,\ldots$, are paired in supermultiplets,
i.e. we have the case of spontaneously broken supersymmetry.
For $\epsilon=+$ we have the case of exact supersymmetry, characterized by
the spectrum
\begin{equation}
E_{n}^{+}=2[(n+1)/2].
\label{ee}
\end{equation}
Here the vacuum state, being a SUSY singlet,
has the energy $E_{0}^{+}=0$,
whereas $E_{n}^{+}=E_{n+1}^{+}>0$ for  $n=2k+1, k=0,1,\ldots$.

Thus, we have
realized the simplest $N=2$ SUSY system
in terms of one bosonic oscillator.
In the coordinate representation, this system
is given by the supercharge operators (\ref{q12})
and by the Hamiltonian
\begin{equation}
H_\epsilon=\frac{1}{2}\left(-\frac{d^2}{dx^2}+x^2 -\epsilon K\right).
\label{hsch}
\end{equation}
The structure of these operators formally is similar to the
structure of corresponding operators in Witten supersymmetric quantum
mechanics \cite{wit} for the simplest system of the
Nicolai superoscillator \cite{nic},
where, in particular, the diagonal Pauli matrix $\sigma_3$
is present in Hamiltonian instead of parity operator $K$.
But this difference turns out to be crucial.
It reveals itself in the fact that the constructed system
contains both phases of exact and spontaneously broken
SUSY, which are distinguished by the parameter $\epsilon$,
whereas in the case of Witten supersymmetric quantum mechanics
both cases $\epsilon=+$ and $\epsilon=-$ give one and the same
superoscillator system with unbroken SUSY. Note also that
in the present system in the phase of exact SUSY ($\epsilon=+$),
the energy level intervals in spectrum (\ref{ee}) are
twice as much as those for the corresponding spectrum of
the superoscillator \cite{nic}.

One can further extend the formal analogy with the simplest Nicolai-Witten
SUSY system. Indeed,
due to the property $E_{n}^{-}>0$, taking place for $\epsilon=-$,
we can construct the Fermi oscillator operators
\begin{equation}
f^{\pm}=\frac{Q^{\mp}_{-}}{\sqrt{H_{-}}}
\label{fdef}
\end{equation}
satisfying the relations
$
\{f^{+},f^{-}\}=1$ and  $f^{\pm 2}=0.$
So, we get a Bose-Fermi (Klein) transformation in terms of one bosonic
oscillator. With the help of these fermionic operators,
satisfying the relation $\{K,f^{\pm}\}=0$,
we can present the hamiltonian $H_{\epsilon}$, given
by eq. (\ref{hexp1}) or (\ref{hsch}),
in the original form of the superoscillator Hamiltonian \cite{nic}:
\begin{equation}
H_{\epsilon}=\frac{1}{2}\{a^{+},a^{-}\}+\epsilon \frac{1}{2}[f^{+},f^{-}].
\label{haf}
\end{equation}
The formal character of such a coincidence can be stressed
once more by the fact of a noncommutativity
of the bosonic creation-annihilation operators and the
fermionic ones,
\begin{equation}
[a^{\pm},f^{\pm}]\neq 0.
\label{af}
\end{equation}
This noncommutativity reveals the essential difference between the
present minimal SUSY bosonization scheme and the nonminimal one,
described in the previous section.
Recall that in the nonminimal scheme the fermionic
operators can be constructed from some bosonic operators
$\tilde{a}{}^{\pm}$ independent from $a^{\pm}$. Therefore, in
the nonminimal bosonization scheme we would have
the Hamiltonian in the same form
(\ref{haf}) but with $[a^\pm,f^\pm]=0$, that would give the
bosonized system exactly corresponding to the Nicolai
superoscillator in the phase of exact supersymmetry.

\nsection{Supersymmetric 2-body Calogero model}
We pass over to the generalizations of the presented simplest
supersymmetric constructions,
and turn to the `$\nu$-deformed' bosonic oscillator system defined by
the deformed Heisenberg algebra \cite{vas1}
\begin{equation}
[a^{-},a^{+}]=1+\nu K.
\label{def}
\end{equation}
Here the Klein operator $K$ is again given by the relations of the form
(\ref{kle}). In the coordinate representation
it can be realized as a parity operator with the help of
eq. (\ref{kpsi}), whereas the deformed creation-annihilation
operators can be realized in the form generalizing
the ordinary case of the undeformed ($\nu=0$) bosonic oscillator
\cite{bri1}--\cite{bri2}:
\begin{equation}
a^{\pm}=\frac{1}{\sqrt{2}}(x\mp ip),
\quad
p=-i\left(\frac{d}{dx}-\frac{\nu}{2x}K\right).
\label{pdef}
\end{equation}
However, it will be more convenient to generalize the previous
constructions in terms of these deformed creation-annihilation
operators and corresponding Fock space, and then to pass over
to the coordinate representation. For the purpose, let us 
introduce the vacuum state and put the sign factor $\kappa=+1$, 
so that $K\vert 0>=\vert 0>.$ We find that the operator 
$a^{+}a^{-}$ acts on the states $|n>=C_n (a^+)^n\vert 0>$ in 
the following way:  $ a^{+}a^{-}|n>=[n]_{\nu}|n>, $ 
$ 
[n]_{\nu}=n+\frac{\nu}{2}(1+(-1)^{n+1}).
$
{}From here we conclude that in the case when
$
\nu >-1,
$
the space of unitary representation of  algebra (\ref{def}), (\ref{kle}) is
given by the complete set of the orthornormal states
$|n>$, in which the corresponding normalization coefficients can be chosen as
$
C_{n}=([n]_{\nu}!)^{-1/2},
$
$[n]_{\nu}!=\prod^{n}_{k=1}[k]_{\nu}$.
Then, proceeding from eq. 
(\ref{def}), one can get the following expression for the
number operator $N$, $N\vert n>=n\vert n>$, in terms of the
operators $a^{\pm}$:
$
N=\frac{1}{2}\{a^{-},a^{+}\}-\frac{1}{2}(\nu+1).
$
Therefore,
we can realize the Klein operator $K$ in terms of the operators
$a^{\pm}$ by means of eq. (\ref{kcos}), and the constructions
carried out with the use of ordinary bosonic oscillator
operators can be repeated here in the same way. So, we get the
supercharges  and Hamiltonian in the form of eqs.  (\ref{qexp})
and (\ref{hexp1}), respectively.  Then, again, we find that
$\epsilon=+$ corresponds to the case of exact supersymmetry.
Here the states $|n>$ are the eigenstates of the Hamiltonian
$H_{+}$ with the same spectrum (\ref{ee}) as in the case of
Heisenberg algebra ($\nu=0$).  On the other hand, for
$\epsilon=-$ we have the case of spontaneously broken
supersymmetry with the shifted energy spectrum:  instead of
(\ref{eb}), we have here
\begin{equation}
E_{n}^{-}=2[n/2]+1+\nu.
\label{scale}
\end{equation}
Hence, in this case the shift of the energy (the scale of the
supersymmetry breaking) is defined by the deformation parameter,
and here we have $E_{n}^{-}>0$ for all $n$
due to the restriction $\nu>-1$, and, therefore,
in the case of the deformed bosonic oscillator we can also realize the
Bose-Fermi transformation with
the help of the relation of the same form (\ref{fdef})
as in the case of the ordinary oscillator.
So, from the point of view of the SUSY constructions, the
deformation of the Heisenberg algebra reveals itself in the scale of
supersymmetry breaking.

Now, let us present the hamiltonian (\ref{hexp1}) in the
coordinate representation with the help of relations
(\ref{pdef}) and (\ref{kpsi}):
\begin{eqnarray}
H_\epsilon&=&
-\frac{1}{2}\left(\frac{d}{dx}+
\left(\epsilon x-\frac{\nu}{2x}\right)K\right)^2
\label{q22}\\
&=&\frac{1}{2}\left(-\frac{d^2}{dx^2}+x^2+\frac{\nu^2}{4x^2}-
\frac{\nu}{2x^2}K-\epsilon(\nu+K)\right).
\label{supcal}
\end{eqnarray}
The expression (\ref{q22}) is a square of the
hermitian supercharge operator $Q^2_\epsilon$
defined by eq. (\ref{q12}), and the system can be interpreted
as a particle minimally coupled to a specific `gauge field' given by
the operator-valued potential $V(x)=i(\epsilon x
-\nu/2x)K$. For $\nu=0$, it
is reduced to the potential $V=i\epsilon xK$, corresponding to
the simplest supersymmetric system considered in the previous
section.  If we omit the last term $\epsilon(\nu+K)$ from eq.
(\ref{supcal}), we reduce the present supersymmetric Hamiltonian
to the Hamiltonian of the 2-body (nonsupersymmetric) Calogero
model \cite{calog} (see refs. \cite{bri1,bri2}).  It is the use
of the deformed Heisenberg algebra that allowed the authors of
refs. \cite{bri1}--\cite{bri2} to simplify considerably the
problem of solving $n$-body Calogero model. So, we see that the
same algebra allows us to get $N=2$ supersymmetric extension of
the 2-body Calogero model without extending the initial system
by fermionic creation-annihilation operators.

\nsection{OSp(2$\vert$2) supersymmetry}
Now we are going to demonstrate that the bosonization constructions
of the $N=2$ SUSY admit the generalization to the case
of more broad OSp(2$\vert 2)$ supersymmetry. This will allow us
to get some further results connected with the 2-body Calogero
model. For the purpose, we note that
the algebra of the OSp(2$\vert$2) supersymmetry contains
$s(2)$ and $osp(1\vert 2)$ superalgebras as subalgebras \cite{osp2}.
Then, using the results
of papers \cite{vas1,mac,ply1} on realization of
$sl(2)$ algebra and more broad $osp(1\vert 2)$ superalgebra
on the Fock space of the {\it deformed} bosonic oscillator,
we construct the following operators:
\begin{eqnarray}
&T_3=\frac{1}{2}(a^{+}a^{-}+a^{-}a^{+}),\quad
T_\pm=\frac{1}{2}(a^{\pm})^{2},
\quad
J=-\frac{1}{2}\epsilon K[a^-,a^+],&
\label{eveng}\\
&Q^\pm =Q^\mp_\epsilon ,\quad
S^\pm=Q^\mp_{-\epsilon}.&
\label{oddg}
\end{eqnarray}
The operators (\ref{eveng}) and (\ref{oddg})
are even and odd generators of OSp(2$\vert$2) supergroup.
Indeed, they form $osp(2|2)$ superalgebra
given by the nontrivial (anti)commutators
\begin{eqnarray}
&[T_3 ,T_\pm ]=\pm 2T_\pm,\quad
[T_- ,T_+ ]=T_3,&\nonumber\\
&\{S^+ ,Q^+ \}=T_+ ,\quad
\{Q^+ ,Q^- \}=T_3 +J,\quad
\{S^+ ,S^- \}=T_3 -J,&\nonumber\\
&[T_+ ,Q^- ]=-S^+ ,\quad
[T_+ , S^- ]=-Q^+,\quad
[T_3 ,Q^+ ]=Q^+,\quad
[T_3, S^- ]=-S^- ,&\nonumber\\
&[J,S^- ]=-S^- ,\quad
[J,Q^+ ]=-Q^+ ,&
\label{osp}
\end{eqnarray}
and by corresponding other nontrivial (anti)commutators
which can be obtained
from eqs. (\ref{osp}) by hermitian conjugation.
Therefore, even generators
(\ref{eveng}) form the subalgebra $sl(2)\times u(1)$, whereas $s(2)$
superalgebra (\ref{salg}), as a subalgebra, is given by the sets
of generators $Q^\pm$ and $T_3+J$, or $S^\pm$ and $T_3-J$.
Moreover, the operators $a^{\pm}$, being
odd generators of $osp(1\vert 2)$
superalgebra (whereas operators $T_3$ and $T_{\pm}$ are its
even generators) \cite{mac},
are expressed in terms of odd generators
of $osp(2|2)$ superalgebra as
$
a^\pm =Q^\pm + S^\pm.
$
Hence, both phases of exact and spontaneously broken
$N=2$ SUSY, discussed above, are contained
in the extended bosonized  OSp(2$|$2) supersymmetry.

Thus, we can conclude that the $osp(2\vert 2)$ superalgebra can
be realized as an operator algebra of the 2-body
(nonsupersymmetric) Calogero model with the help of the deformed
Heisenberg algebra involving the Klein (parity) operator $K$.
Moreover, the given construction means that the OSp(2$\vert$2)
supersymmetry is a dynamical symmetry for the bosonized
supersymmetric extension of the 2-body Calogero model presented
in the previous section.

Note that the supersymmetric extension of the $n$-body Calogero
model \cite{calog} was realized in ref. \cite{freed} in a
standard way by introducing fermionic degrees of freedom into
initial nonsupersymmetric system.  In ref. \cite{bri2}, the
supersymmetric extension of the $n$-body Calogero model was
investigated with the help of the $n$-extended deformed Heisenberg
algebra supplied with the corresponding set of the fermionic
creation-annihilation operators, where OSp(2$\vert$2)
supersymmetry was also revealed as a dynamical symmetry of the model
of ref. \cite{freed}.  Therefore, the constructions given here
bosonize corresponding SUSY constructions of ref. \cite{bri2}
for the 2-body case.

\nsection{Bosonized supersymmetric quantum mechanics}
We turn now to the generalization of the
previous bosonization constructions
of the $N=2$ SUSY to the case
corresponding to the more complicated
quantum mechanical $N=2$ supersymmetric
systems \cite{wit,salhol,gen}. To this end, consider the operators
$
\tilde{Q}^{\pm}_{\epsilon}=A^{\mp}\Pi_{\pm\epsilon}
$
with mutually conjugate odd operators $A^{\pm}=A^{\pm}(a^{+},a^{-})$,
$A^{-}=(A^{+})^{\dagger}$, $KA^{\pm}=-A^{\pm}K$.
These properties of $A^{\pm}$ guarantee that the
operators
$\tilde{Q}^{\pm}_{\epsilon}$ are, in turn, mutually conjugate,
$\tilde{Q}^{-}_{\epsilon}=
(\tilde{Q}^{+}_{\epsilon})^{\dagger}$, and nilpotent:
$
(\tilde{Q}{}^{\pm}_{\epsilon})^{2}=0.
$
Taking the anticommutator
$
\tilde{H}_{\epsilon}=\{\tilde{Q}^{+}_{\epsilon},\tilde{Q}^{-}_{\epsilon}\}
$
as the Hamiltonian,
we get the  $N=2$ superalgebra with the generators
$\tilde{Q}_\epsilon$ and $\tilde{H}_\epsilon$.
The explicit form of the supersymmetric Hamiltonian 
$\tilde{H}_\epsilon$ has the form given by eq. (\ref{hexp1}) 
with operators $a^\pm$ replaced by $A^\pm$.  Now, let us turn 
to the coordinate representation, and choose the operators 
$A^{\pm}$ in the form 
$A^{\pm}=\frac{1}{\sqrt{2}}(\mp ip+\tilde{W}(x,K))$ 
with odd function  $\tilde{W}(x,K)$, 
$K\tilde{W}(x,K)=-\tilde{W}(x,K)K$,
which generally can depend on the parity operator $K$,
and, so, has the form $\tilde{W}=W_0(x)+iW_1(x)K$,
where $W_0(x)$ and $W_1(x)$ are real odd functions.
We shall call $\tilde{W}$ a superpotential.
Taking into account realization (\ref{pdef}) for the
deformed momentum operator $p$, as a result we get the following
most general form of the $N=2$ supersymmetric Hamiltonian,
quadratic in the derivative $d/dx$,
\begin{equation}
\tilde{H}_{\epsilon}=
-\frac{1}{2}\left(\frac{d}{dx}+
i\epsilon W_1 +\left(\epsilon W_0-\frac{\nu}{2x}
\right)K\right)^2.
\label{swit}
\end{equation}
Here the Hamiltonian is written formally as a square
of the corresponding supercharge operator
$\tilde{Q}{}^2_\epsilon =i(\tilde{Q}{}^-_\epsilon
-\tilde{Q}{}^+_\epsilon)$.
Therefore, from the point
of view of the present constructions,
the $N=2$ supersymmetric system given by
the superpotential $\tilde{W}=W_0+iW_1 K$
in the case of
the deformed Heisenberg algebra
(\ref{def}), (\ref{kle}) is equivalent
to supersymmetric system given
by the shifted superpotential
$\tilde{W}=(W_0-\epsilon \nu/2x)+iW_1 K$
in the undeformed  case ($\nu=0$).
In particular, in terms of the
ordinary ($\nu=0$)  Heisenberg algebra,
the $N=2$ supersymmetric extension of the 2-body
Calogero model, constructed in section 4, is
the supersymmetric system given by the superpotential
with $W_1=0$ and $W_0=x-\epsilon \nu/2x$.
Moreover, as follows from the explicit
form of the supersymmetric Hamiltonian (\ref{swit}),
the function $W_1$ can be eliminated from the superpotential
by the phase transformation of a wave function,
$\psi(x)\rightarrow \tilde{\psi}(x)=
\exp(-i\epsilon\int_{}^{x}W_1(x')dx')\psi(x)$.
Therefore, finally we arrive at the
following general form of the supersymmetric
Hamiltonian and corresponding selfconjugate supercharge operators,
\begin{eqnarray}
&&H=\frac{1}{2}\left(-\frac{d^2}{dx^2}+W^2(x)- W'\cdot K\right),
\nonumber\\
&&Q_1=iQ_2\cdot K,\quad
Q_2=-\frac{i}{\sqrt{2}}\left(\frac{d}{dx}+
W(x)\cdot K\right),
\label{swit0}
\end{eqnarray}
which are defined by {\it odd function} $W(x)$ being a superpotential.
This formally corresponds to the case of
Witten supersymmetric quantum mechanics \cite{wit}, in which
$N=2$ supersymmetric system is also
defined by one function being a superpotential.
However, it is necessary to stress once more
that here the superpotential is an odd
function, and, as it has been shown
with the help of the simplest
system given by a linear superpotential
$W=\epsilon x$, the present construction
relates to the Witten supersymmetric quantum mechanics
in a nontrivial way.

\nsection{Concluding remarks and outlook}
When the operator $K$ is understood as the Klein operator given
by eq. (\ref{kcos}) in terms of creation-annihilation operators, 
the described bosonization of supersymmetric quantum mechanics
is minimal in the sense that it is realized on the Fock space of
one (ordinary or deformed) bosonic oscillator.  On the other
hand, in the coordinate representation the operator $K$ can be
considered as a parity operator defined by relation
(\ref{kpsi}).  We have illustrated a nontrivial relationship of
the construction to the Witten supersymmetric quantum mechanics
with the help of the simplest $N=2$ SUSY system, and have shown
that the essential difference between the minimal SUSY
bosonization scheme and the nonminimal one, discussed at the
beginning of the paper, is coded in relation (\ref{af}).  The
general case of the bosonized Witten supersymmetric quantum
mechanics is given by the Hamiltonian and supercharges
(\ref{swit0}), which, in turn, are defined by odd
superpotential. So, it would be interesting to investigate the
general properties of the bosonized $N=2$ SUSY and establish its
exact relationship to the Witten supersymmetric quantum
mechanics.

We have revealed the OSp(2$\vert$2) supersymmetry in the
system being the bosonized supersymmetric 2-body Calogero model.
The open problem here is establishing the criteria for the existence of
such an extension of the $N=2$ supersymmetry in general case
of the bosonized Witten supersymmetric quantum mechanics (\ref{swit0}).

The classical analog of the Witten
supersymmetric quantum mechanics is formulated on the
superspace containing Grassmann variables being the classical
analogs of the fermionic operators, and the corresponding
path-integral formulation of the theory is well known (see,
e.g., ref. \cite{das}).
What is the classical analog
and corresponding path-integral formulation
for the bosonized supersymmetric system given by eq.
(\ref{swit0})?  This question is very interesting because a
priori it is not clear at all how the supersymmetry will reveal
itself without using Grassmann variables (in this respect see,
however, ref.  \cite{pls}).  Possibly, the answer can be
obtained using recent result on realization of the classical
analog of the $q$-deformed oscillator with the help of
constrained systems \cite{sha} and the known observation on the
common structure of different bosonic deformed systems
\cite{dasc}.

As a further generalization of the constructions, one could
investigate a possibility to bosonize S(N)-supersymmetric, $N>2$,
\cite{cromb}  and parasupersymmetric \cite{para}
quantum mechanical systems.

Another obvious development of the present bosonization
constructions would be their generalization to the case of $n>1$
bosonic degrees of freedom.  Possibly, in this direction
$osp(2\vert 2)$ superalgebra could be revealed in the form of
the operator algebra for the general case of the $(n+1)$-body
nonsupersymmetric Calogero model, and, moreover, a
supersymmetric extension of this model could be constructed
without supplying the system with fermionic degrees of freedom.

Then, taking the limit $n\rightarrow\infty$, one could try to
generalize SUSY bosonization constructions to the case of
(1+1)-dimensional quantum field theory.  In connection with such
hypothetical possible generalization it is necessary to point
out that earlier some different problem was investigated by
Aratyn and Damgaard \cite{damg}.  They started from the
(1+1)-dimensional supersymmetric field system containing free
complex scalar and Dirac fields, and bosonized fermionic field
in terms of an independent scalar field with the help of the
Mandelstam nonlocal constructions \cite{bosf}, i.e. used
nonminimal bosonization scheme according to our terminology.
As a result, they arrived at the quantum field system of free
bosonic scalar fields, described by a local action.  On the
other hand, due to relation (\ref{af}) one can conjecture that
the corresponding quantum field generalization of the present
minimal SUSY bosonization constructions will give some different
results.

At last, it seems to be interesting to investigate the
possibility of realizing the bosonized supersymmetric extension
of (2+1)-dimensional anyonic equations constructed in ref.
\cite{ply1} with the help of the deformed Heisenberg algebra.

We hope to consider the problems listed here in future publications.

$\ $

The author thanks P.H. Damgaard, T.H. Hansson,
A. Niemi and P.Di~Vecchia for useful discussions.
The research was supported  by MEC-DGICYT (Spain).


\end{document}